\documentclass[]{aastex631}
\usepackage{amsmath}

\newcommand{\vishal}[1]{\textcolor{black}{#1}}

\newcommand{\car}{\ion{C}{2}~}
\newcommand{\mg}{\ion{Mg}{2}~}
\newcommand{\bmag}{$\vert$B$\vert$}
\shorttitle{Properties of \rm{C}{2} line}
\shortauthors{Upendran \& Tripathi}
\graphicspath{{./}{figures/}}

\begin{document}
\nolinenumbers
\title{Properties of the \car 1334~{\AA} line in Coronal Hole and Quiet Sun as observed by IRIS}
\correspondingauthor{Vishal Upendran}
\email{uvishal@iucaa.in}

\author[0000-0002-9253-6093]{Vishal Upendran}
\author[0000-0003-1689-6254]{Durgesh Tripathi}
\affiliation{Inter-University Centre for Astronomy and Astrophysics, Post Bag-4, Ganeshkhind, Pune 411007, India)}

\begin{abstract}
Coronal Holes (CHs) have subdued intensity and net blueshifts when compared to Quiet Sun (QS) at coronal temperatures. At transition region temperatures, such differences are obtained for regions with identical photospheric absolute magnetic flux density ({\bmag}). In this work, we use spectroscopic measurements of the \car 1334~{\AA} line from Interface Region Imaging Spectrograph (IRIS), formed at chromospheric temperatures, to investigate the intensity, Doppler shift, line width, skew, and excess kurtosis variations with {\bmag}. We find the intensity, Doppler shift, and line widths to increase with {\bmag} for CHs and QS. The CHs show deficit in intensity and excess total widths over QS for regions with identical {\bmag}. For pixels with only upflows, CHs show excess upflows over QS, while for pixels with only downflows, CHs show excess downflows over QS that cease to exist at {\bmag} $\le$ 40. Finally, the spectral profiles are found to be more skewed and flatter than a Gaussian, with no difference between CH and QS. These results are important in understanding the heating of the atmosphere in CH and QS, including solar wind formation, and provide further constraints on the modeling of the solar atmosphere.
\end{abstract}

\section{Introduction} \label{sec:intro}

The upper solar atmosphere, i.e., solar corona, is highly structured, and may be broadly classified into the bright Active Regions (ARs), the dark Coronal Holes (CHs), and the remaining areas with no extensive activity called Quiet Sun (QS). While all these regions can be visually differentiated in the corona, the visual distinction between CHs and QS cannot be made in the intensities recorded at lower atmospheric heights, such as the transition region and the chromosphere.

Various investigators have undertaken comprehensive studies of CH and QS across different temperatures~\citep[see e.g.][]{Stucki_FUVLinesSumer,Stucki_UVLinesSumer_Chs,xia_chsumer}, and studied the line intensity, velocity and widths covering temperatures from $\approx8\times10^3$ K to  $\approx1.4\times10^6$ K. It was found that CHs have marginally excess intensity over QS for the spectral lines with peak formation temperature between $\approx8\times10^3$ K to $\approx1.42\times10^5$ K. But, the spectral lines forming at higher temperatures showed excess intensity in QS over CH. Furthermore, CHs were found to be redshifted (blueshifted) with respect to QS for lines forming at temperatures below(above) $\approx6.76\times10^4$ K. Finally, CHs were seen to have larger line widths than QS for almost all the lines in observation. However, note that the differences in intensities, velocities and widths in the lower atmosphere were negligible within the errors \citep[see e.g][for details]{Stucki_FUVLinesSumer}.

The \ion{He}{1} 10830~{\AA} and \ion{He}{1} 584~{\AA} lines are the only chromospheric lines, which show significant differences between CH and QS. Note that the \ion{He}{1}~10830~{\AA} is an absorption line, while the \ion{He}{1}~584~{\AA} is an emission line. The \ion{He}{1}~10830~{\AA} line shows excess intensity in CHs \citep{harvery_heI10830,Kahler_heI10830}, while the \ion{He}{1} 584~{\AA} line shows lower intensity in CHs \citep{Jordan_heI584} over QS. The \ion{He}{1} 584~{\AA} was also found to have excess blueshifts and linewidths inside CHs, when compared to QS regions~\citep{Peter_1999_HeICHQS}. However, these differences may be attributed to the sensitivity of these lines to the EUV radiation from the corona, which provides coronal information in the chromosphere. At radio wavelengths (1.21 mm), CHs are indistinguishable from QS~\citep{brajvsa2018_CHradio}. This however, is not the case in microwave. At 17~GHz, CHs are found to be brighter than QS, with a variability over multiple time scales~\citep{gopalswamy1999_CHmicrowave}. This variability, and the dynamic nature of CHs is suggested by \cite{gopalswamy1999_CHmicrowave} to be a signature of solar wind acceleration and heating.

Differences between CHs and QS have also been studied in chromosphere and transition region using the \ion{Mg}{2} \citep{PradeepKashyap2018,bryans_swchrom} and \ion{Si}{4} lines \citep{TriNS_2020} recorded by the Interface Region Imaging Spectrograph \citep[IRIS,][]{iris}. \cite{PradeepKashyap2018} found that while there is no distinction in the distribution of the absolute photospheric magnetic flux density ({\bmag}) in CH and QS, the QS shows excess intensity vis-\`a-vis CHs for regions with identical {\bmag}. \cite{bryans_swchrom} found that some CHs may be identified through a marginal excess in the peak separation of the \ion{Mg}{2}~h line. Similarly, for \ion{Si}{4} line, \cite{TriNS_2020} found that CHs and QS show differences in intensities and Doppler shifts for the regions with identical {\bmag}, while no difference in the non-thermal width was observed.

These observations lead us to the question -- are CHs and QS already differentiated at the chromosphere? If so, how are plasma properties different in CHs and QS at this level? Moreover, how do these properties vary with the underlying magnetic field? A detailed understanding of these questions will not only provide essential ingredients to further understand the heating and dynamic coupling of the solar atmosphere in CHs and QS, but also help us diagnose the origin \& formation of the solar wind \citep[see e.g.,][]{HassDL_1999,tu2005solar} and the recently observed switch-backs in the magnetic field \citep[see e.g.,][]{balogh_1999_switchback,fisk_2003_switchbacks,Bale_2019_switchbacke,fisk_2020_switchbacks,Zank_2020_theory,TriNS_2020}.

\vishal{For the above described purpose,  we study the properties of the \car 1334~{\AA} line observed by IRIS in CHs and QS as a function of {\bmag}. The remaining paper is structured as follows: In \S \ref{sec:Data}, we present details of our observations, with description of data preprocessing in \S\ref{sec:dataprepocess}, feature extraction in \S\ref{sec:featureextract}, and segmentation in \S\ref{sec:segmentation}. The obtained results are presented in \S\ref{sec:results}. We finally summarize, discus and conclude in \S\ref{sec:discuss}.}


\section{Observations and Data}\label{sec:Data}
In this work, we have primarily used the observations recorded from IRIS, along with those from the Atmospheric Imaging Assembly \citep[AIA;][]{BoeEL_2012} and Helioseismic and Magnetic Imager~\citep[HMI;][]{HMI}, on board the Solar Dynamics Observatory~\citep[SDO;][]{SDO}. IRIS observes the Sun in three wavelength bands, \textit{viz.} from 2782.7 to 2851.1~{\AA} in near ultraviolet (NUV), from 1332 to 1358~{\AA} in far ultraviolet-1 (FUV 1), and from 1389 to 1407~{\AA} (FUV 2). We use the \car rasters, which have a pixel size of $\approx0.16${\arcsec} along the slit, and sample at $\approx.35${\arcsec} across the field of view (FOV), along with a spectral pixel size of $\approx25.9$m{\AA}. Time time cadence between successive slit positions is $\approx30$ seconds. 

IRIS also has a Slit-Jaw Imager, which provides photometric Slit-Jaw Images (SJIs) with passbands around the strong lines in NUV and FUV centered around 1330~{\AA} and 2796~{\AA}. We use the SJIs for co-alignment purposes. These SJIs have a pixel size of $\approx0.16${\arcsec}, and are available at a cadence of $\approx63$ seconds. 

AIA observes the Sun's atmosphere in UV and EUV bands using eight different passbands sensitive to plasma at different temperatures \citep{ODwyer, BoeEL_2012}. Here, we use data from the 193~{\AA} and 1600~{\AA} passband from AIA, and the Line-of-sight (LOS) magnetic field data from HMI. The AIA images are taken with a pixel size of $\approx$0.6{\arcsec} and a time cadence of $\approx12$~s. The LOS magnetic field measurements ($B_{LOS}$) are obtained by HMI at $\approx45$~s cadence with a pixel size of 0.5{\arcsec}. \vishal{Note that we do not use the vector magnetogram measurements from HMI due to larger noise in this product, especially for CHs and QS regions~\citep{Hoeksema_2014_HMIpipeline,couvidat2016observables}}.We have used images taken at 193~{\AA} to identify CHs and QS and those at 1600~{\AA} for co-alignment between IRIS and HMI.
\begin{deluxetable*}{|c|c|c|c|c|}
  \tablecaption{Details of the IRIS rasters used in this study. \label{tab:datadetails}}
  \tablewidth{0pt}
  \tablehead{\colhead{Dataset name} & \colhead{Time range} & \colhead{(Xcen,Ycen)} & \colhead{Raster FOV} & \colhead{$\mu$}}
   \startdata
  DS1 & 2014-07-24 11:10:28 -- 14:40:53 & (128{\arcsec},-180{\arcsec}) & (141{\arcsec},174{\arcsec}) & 0.97 \\
  DS2 & 2014-07-26 00:10:28 -- 03:40:53 & (469{\arcsec},-167{\arcsec}) & (141{\arcsec},174{\arcsec}) & 0.85\\
  DS3 & 2014-08-02 23:55:28 -- 03:25:53 +1d & (332{\arcsec},-152{\arcsec}) & (141{\arcsec},174{\arcsec}) & 0.92\\
  DS4 & 2015-04-26 11:39:31 -- 15:09:56 & (-288{\arcsec},45{\arcsec}) & (141{\arcsec},174{\arcsec}) & 0.95\\
  DS5 & 2015-10-14 11:07:33 -- 14:37:58 & (215{\arcsec},-165{\arcsec}) & (141{\arcsec},174{\arcsec}) & 0.97\\
  \enddata
\end{deluxetable*}

For our study, we considered five different IRIS rasters obtained during a time span of a year. We have selected the observations such that the QS and CHs are present in the same FOV of each raster. The details of the IRIS observations are given in Table.~\ref{tab:datadetails}. Note that DS1, DS2 and DS5 are the same ones used by \cite{TriNS_2020}. We use corresponding coordinated AIA data cubes for this study, while the full disk HMI cubes were cut out from Level{--}2 data. 

\begin{figure}[ht!]
  \includegraphics[width=\linewidth]{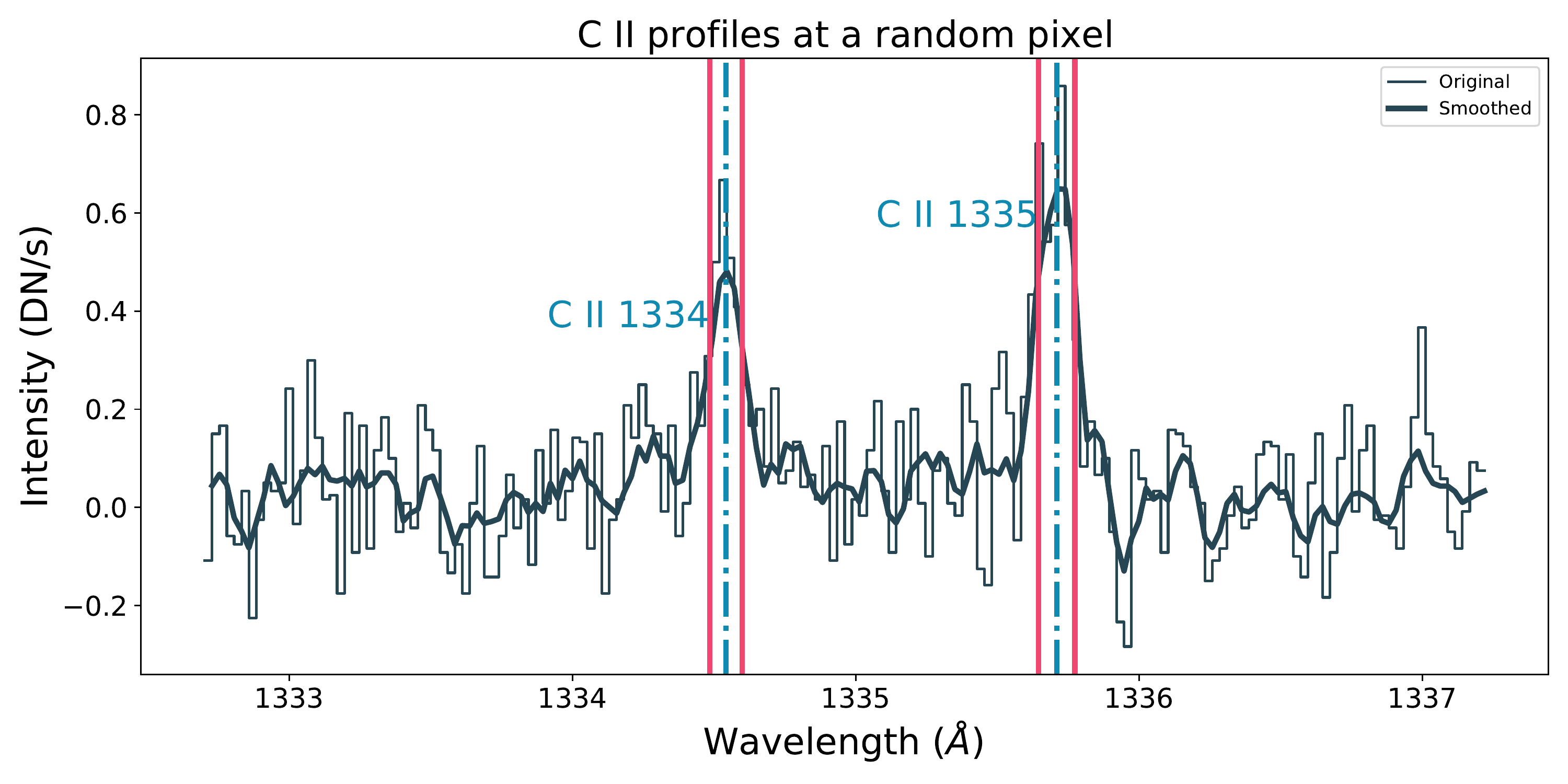}
  \caption{An example spectrum centered at the two \car lines obtained at a random QS location of DS3. The step plot shows the original spectrum, while the black solid line is the smoothed locally-averaged spectrum (as described in \S\ref{sec:dataprepocess}). The two \car lines are marked, with the dot-dashed blue lines depicting the line center, and the red solid lines depicting $\pm\sigma$, both obtained from a single Gaussian fit to each line. }
  \label{fig:rasterfov}
\end{figure}

The \car lines \vishal{at 1334~{\AA} and 1335~{\AA}} form in an optically thick atmosphere under non-LTE conditions. Hence, depending on the variation of source function with height, the profiles can show single-peaked, double peaked or even multi-peaked profiles \citep{Rathore_CII_paper1}, though the profiles are predominantly single peaked in QS regions \citep{Rathore_CII_paper3}. \vishal{Note however that on an average, the line centroid is well correlated with the atmospheric velocity at line core formation height~\citep[see Fig. 15 of][]{Rathore_CII_paper2}. The line Full Width at Half Maximum (FWHM) deviates marginally from the theoretical result expected from a single Gaussian fit~\citep[see Fig. 16.a of][]{Rathore_CII_paper2}. However, since we are interested in comparing the line widths across CH and QS, the absolute values in themselves may be underestimated by similar factors in both these regions. 
Similarly, we seek to compare the intensities across CH and QS, and hence a single Gaussian fit is justified for such a comparison. Nevertheless, we also seek to understand the deviations of profiles from a single Gaussian for both CHs and QS. Hence, we also study the higher moments of the \car line in CH and QS. Note however that given {\bmag} is not very different between CH and QS,  we expect the \car line profiles to show similar behavior in CH and QS. }

An example QS spectrum centered at the two \car lines obtained from a random pixel of DS3 is shown in Fig.~\ref{fig:rasterfov}. The two \car lines, along with their centroids (blue dot-dashed) and $\pm\sigma$ locations (solid red) as obtained from a single Gaussian fit, are marked with vertical lines. Note that the $1335.707$~{\AA} line is actually blended with another \car line at $1335.66${~\AA}. \vishal{Hence, we do not use the 1335 line in our analysis, due to ambiguity introduced in the analysis by the line blend. To estimate the gross properties of the 1334 line, we hence perform a single Gaussian fit and extract the parameters out.}  

\subsection{Data pre-processing} \label{sec:dataprepocess}

Since AIA, HMI and IRIS are distinct instruments, we perform a co-alignment between these observations before performing detailed analysis. For this purpose, first, the SJIs obtained from IRIS, and $B_{LOS}$ maps from HMI are down-sampled to AIA resolution. The AIA 1600~{\AA} images are then co-aligned with the nearest SJIs from IRIS at 2796~{\AA}. The shifts from co-alignment are then applied to AIA 193~{\AA} and HMI $B_{LOS}$ data. Next, we generate artificial rasters of AIA and $B_{LOS}$ data by selecting data along the slit position from these images, for the observation time. Finally, both the AIA and HMI pseudo-rasters are converted to IRIS raster FOV and  resolution to enable comparison with features derived later on from these rasters. \vishal{Note that we use the \texttt{sunpy.map.resample} function to make the resolutions uniform, which uses linear interpolation under the hood.}

\subsection{Feature extraction and spectral properties} \label{sec:featureextract}

We first smooth the spectral profiles following \cite{Rathore_CII_paper3}. This smoothing marginally increases the number of converged fits, especially in regions with low intensity. The smoothing filter taken from \cite{Rathore_CII_paper3} is:

\begin{equation*}
  S_{\text{filt}} = \begin{cases}
  \frac{\sigma^2}{\sigma_s^2}m_s+\left(1-\frac{\sigma^2}{\sigma_s^2}\right)s & \text{$\sigma_s^2 \ge \sigma^2$}\\
  m_s & \text{$\sigma_s^2$ $<$ $\sigma^2$}
  \end{cases}
  \label{eqn:rpcd}
\end{equation*}

\noindent where the $S_{\text{filt}}$ is the filtered signal in a 3$\times$3 window, where $m_s$ and $\sigma_s^2$ are the local means and variances, while $\sigma^2$ is average of local variances. For regions with strong signal, the S$_{\text{filt}}$ tends to the local mean $m_s$, while the weaker regions are smoothed out. This operation is performed in slices of 2-D spectrogram of [coordinate along the slit, wavelength]. On the obtained spectra, we perform a single Gaussian fit with a constant continuum to the \car line profiles following \cite{Rathore_CII_paper2}. This scheme, while having the disadvantage of being influenced by whole line profile in providing line core information, was our best bet due to the relatively large noise in using a peak finding algorithm. 

The fit is performed within a spectral window of $\pm50$~km~s$^{-1}$ with respect to the reference wavelength of $1334.532$~{\AA}, as taken from \cite{Rathore_CII_paper1,kelly_c2linerefwave}. From this fitting, we obtain the line core intensity, Doppler shift (i.e, the centroid) and width. The smoothed spectrum (solid) and the fitted line centroid, $\sigma$ are depicted in Fig.~\ref{fig:rasterfov}. As mentioned earlier, the line core intensity is a proxy for strength of the source function, as shown in ~\cite{Rathore_CII_paper2}. Similarly, the Doppler shift is a measure of the plasma velocity at the formation height. The line width, however, is a function of both the line formation temperature and opacity broadening factor, as shown in \cite{Rathore_CII_paper2}. Double-peaked profiles are formed due to the presence of a local maximum in the source function, while line profiles become asymmetric due to the presence of velocity gradient in the chromosphere. For further information on the formation of \car lines and their general properties see \cite{Rathore_CII_paper2,Rathore_CII_paper3,Avrett_2013_1DC2}.

We also estimate the third and fourth moments namely the skew and kurtosis, respectively, of the spectral profiles following \cite{jeffrey2016moments}. These are computed since the observed spectral profiles are known to have marked departures from a  Gaussian profile~\citep{Rathore_CII_paper3}. The skew and kurtosis for a perfectly Gaussian profile are expected to be 0 and 3, respectively. Hence any departures would indicate a significant difference from a Gaussian profile. 

The skew (S) and the excess kurtosis (K) are defined as: 
\begin{equation}
  S = \frac{1}{\sigma^3}\frac{\int_{\lambda} I(\lambda)~(\lambda-\lambda_D)^3 d\lambda}{\int_{\lambda}I(\lambda)~d\lambda},
  \label{eqn:skew}
\end{equation}
\begin{equation}
  K = \frac{1}{\sigma^4}\frac{\int_{\lambda} I(\lambda)~(\lambda-\lambda_D)^4 d\lambda}{\int_{\lambda}I(\lambda)~d\lambda} - 3.0,
  \label{eqn:kurt}
\end{equation}

\noindent where $\lambda_D$ is the centroid estimated from the Gaussian fits, and the integral is performed over the range $\pm50$ km/s of our spectral window in wavelengths, around the reference wavelength. The $\sigma^2$ is the second moment of the line given by

\begin{equation}
  \sigma^2 = \frac{\int_{\lambda} I(\lambda)~(\lambda-\lambda_D)^2 d\lambda}{\int_{\lambda}I(\lambda)~d\lambda}.
  \label{eqn:var}
\end{equation}

Note that the moments are computed for the Gaussian fit to the line. For the spectral line, the continuum is subtracted, and then the moments are computed, following~\citet{jeffrey2016moments}

\subsection{Segmentation of CH and QS} \label{sec:segmentation}
Since our is aim to study the properties of \car line, both in CHs and QS, we generate segmentation maps separating CHs and QS. To isolate the CHs and QS, previous investigators \citep[e.g.,][]{PradeepKashyap2018, TriNS_2020} have considered an intensity threshold of 80 DN in AIA 193~{\AA} images. While such an intensity threshold does a good job, it is an ad-hoc procedure. To be more objective and have an adaptive threshold depending on the distribution of intensity, we follow the threshold method outlined in \cite{Upendran_solarwind}, which is based on Otsu's algorithm \citep{Otsu}. The algorithm works by assuming the presence of two distinct distributions of pixel intensities separated by maximizing the inter-class variance. \cite{Upendran_solarwind} modified this algorithm by applying a `stacked thresholding' to separate out the CH and QS clearly. 

\section{Data analysis and Results}\label{sec:results}

As stated above, we have studied five different IRIS rasters that contained CHs and QS within the same FOV. We study the dependence of spectral line properties {\it viz.} intensity, Doppler shift and width on {\bmag} in CH and QS. We performed identical analysis on all five data sets. Here, we discuss in detail the results for DS3. Note that similar results were obtained for other four data sets. Since all the observations are taken at similar $\mu$-values on the solar disk, to improve the statistics, we combine the results from all data sets in \S\ref{sec:alldatasets} by averaging the obtained parameters from different data sets. Note that we consider 10 Gauss as the noise in the magnetic field density \citep{yeo_10gauss_bfielderror,couvidat2016observables}. 

To improve Signal to Noise Ratio (SNR) and statistics, we consider the derived quantities in bins of {\bmag}, and report the average values in these bins. For this purpose, we use a constant {\bmag} bin size of 0.1 in log-space to account for the lesser number of pixels at higher {\bmag}. Note that the LOS values have been converted to radial values for both {\bmag} \& Doppler shifts via division by $\mu$. Note also that the error bars reported in all the plots are the standard error on the mean. The standard error is defined as $\sigma/\sqrt{N}$, where $\sigma$ is the standard deviation of the samples present in each bin, and $N$ is the number of points in the bin. This error quantifies uncertainty in estimating the mean value of a quantity.

\begin{figure}[htpb!]
\centering
\includegraphics[width=\linewidth]{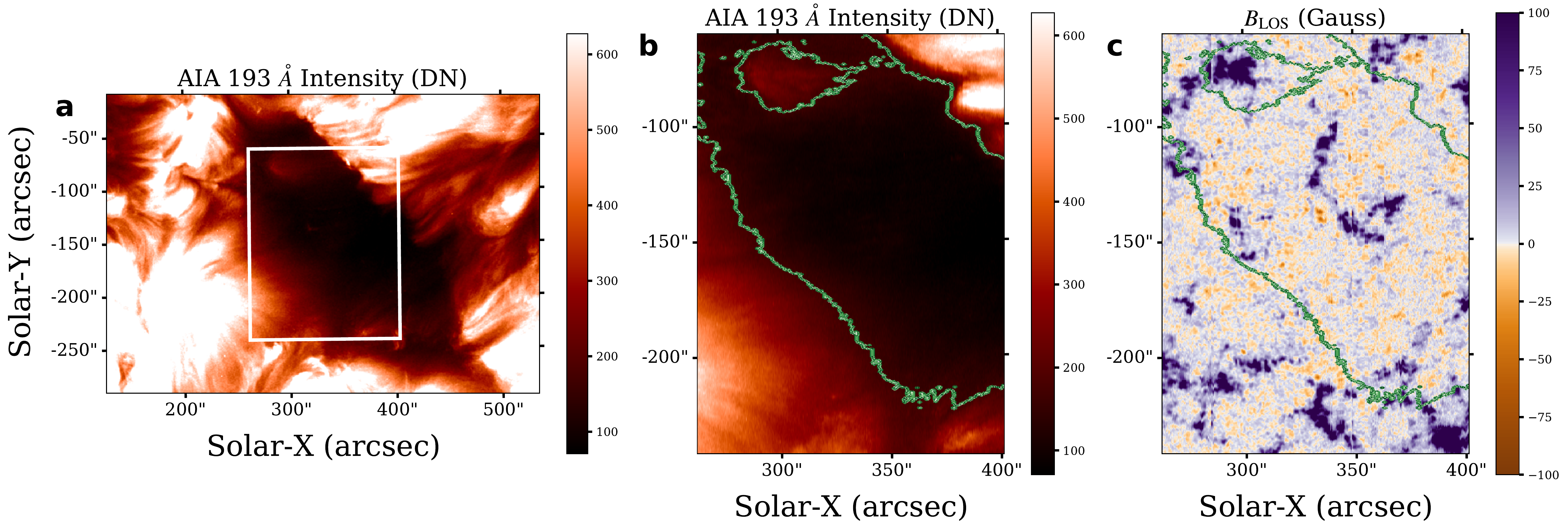}
\caption{Context image and pseudo-rasters from DS3. Context image from AIA 193~{\AA} is shown in  panel~\textbf{a}. The over-plotted white box shows the IRIS raster field of view. Panels~\textbf{b} and  \textbf{c} show the pseudo-rasters obtained from AIA 193~{\AA} and HMI LOS magnetograms, respectively. The green contours in panels \textbf{b} and \textbf{c} depict the boundary between the CH and QS obtained from the segmentation algorithm as described in \S\ref{sec:segmentation}.} \label{fig:segmentation_profiles}
\end{figure}

In Fig.~\ref{fig:segmentation_profiles}.a, we display the context AIA 193~{\AA} image. The over-plotted white box corresponds to the IRIS raster FOV. Fig.~\ref{fig:segmentation_profiles}.b \& c, displays the pseudo-rasters of AIA 193~{\AA} and the line of sight (LOS) magnetic field map, respectively. The over-plotted green contours in panels b and c demarcate the boundary between the CH and QS. While there is clear difference between the CH and QS is observed in the AIA image, no such difference is seen in the magnetograms.

\subsection{Line Intensities} \label{sec:c2intensity}
\begin{figure}[!ht]
\centering 
\includegraphics[width=0.66\linewidth]{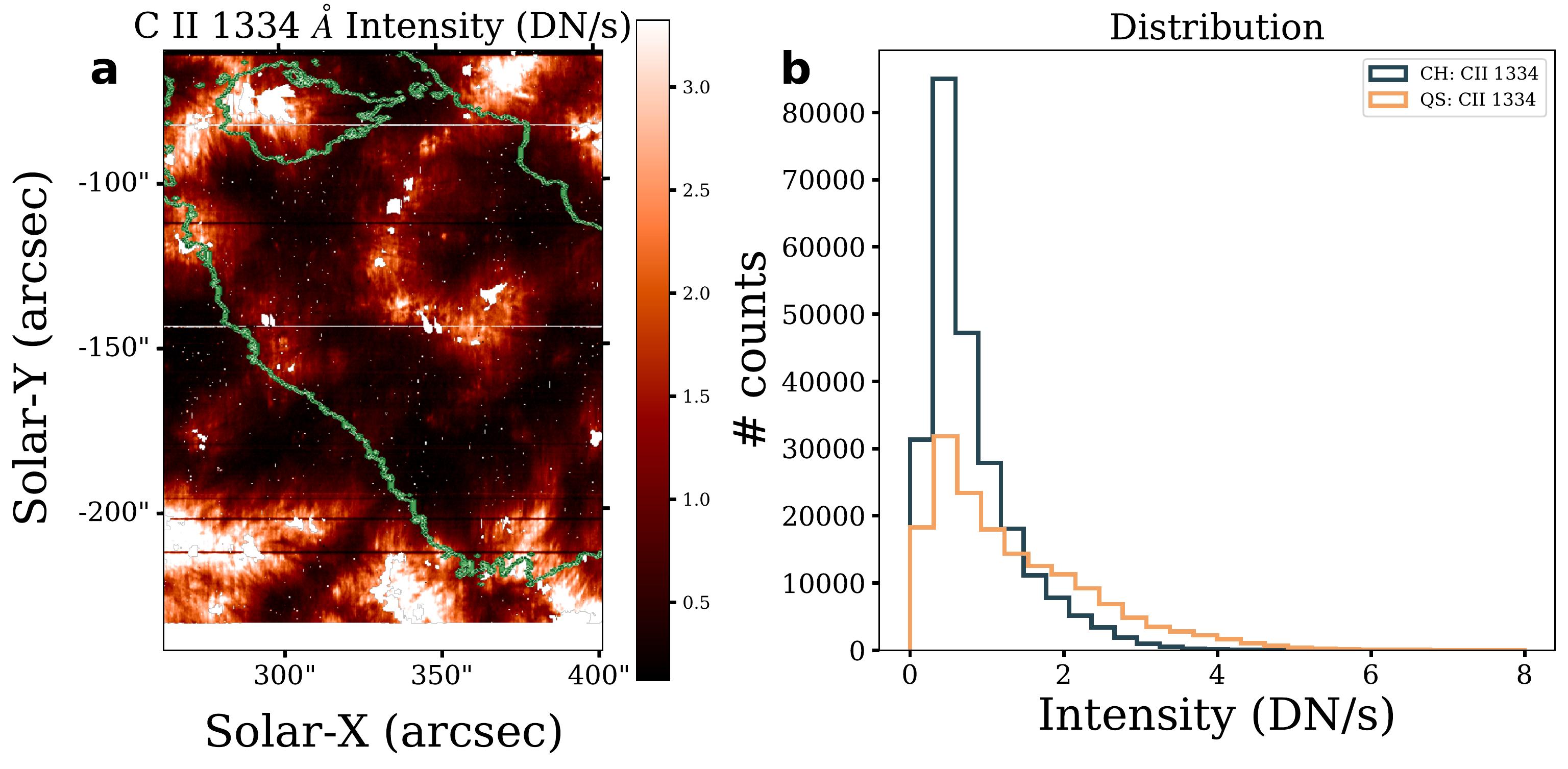}
\includegraphics[width=0.33\linewidth]{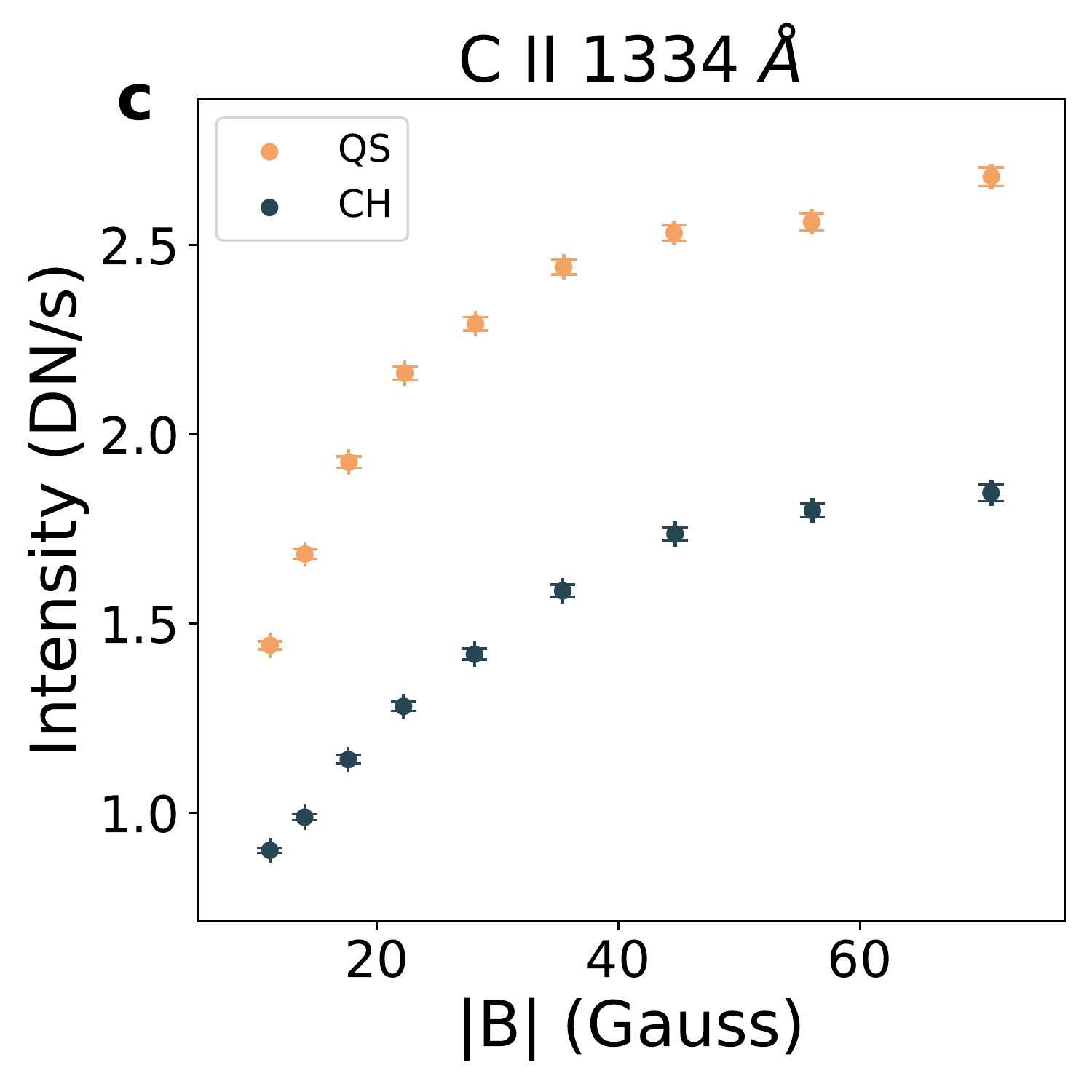}
\caption{ \vishal{Intensity map of \car 1334~{\AA} for DS3 is shown in panel \textbf{a}.} The green contours show the boundary between the between CH and QS. The intensity distribution in CH (black) and QS (orange) are shown in panel \textbf{b}. \vishal{The \car 1334~{\AA} intensity variation with {\bmag} for DS3 is shown in panel \textbf{c}. The orange color indicates QS, and black indicates CH.} }
\label{fig:c2i_map}
\end{figure}

\vishal{Fig.~\ref{fig:c2i_map}.\textbf{a} displays the intensity map obtained in \car 1334~{\AA}.} The over-plotted green contours are the same as those plotted in Fig.~\ref{fig:segmentation_profiles}.b. Note that the intensity map, and all subsequent maps show a white space at the bottom of the raster that corresponds to missing data. \vishal{In Fig.~\ref{fig:c2i_map}.\textbf{b}, we plot the intensity distribution obtained within the CH (black) and QS (orange).} The number of bins of the histogram is selected using Doanne's rule \citep{doanerule}. There is no visual difference between the CH and QS in Fig.~\ref{fig:c2i_map}. a like the differences seen in the coronal image of Fig.~\ref{fig:segmentation_profiles}. \vishal{Note however that excess counts are seen at higher intensities for QS over CH in the distribution in Fig.~\ref{fig:c2i_map}. b.} 

From the intensity maps shown in Fig.~\ref{fig:c2i_map}.a, and the photospheric magnetic field maps shown in Fig.~\ref{fig:segmentation_profiles}.c, we find a clear correspondence between the {\bmag} and intensities. In \vishal{Fig.~\ref{fig:c2i_map}.c, we plot the variation of intensities as a function of {\bmag}}. We find that for both CHs and QS, the intensities of the \car line increases with increasing {\bmag} till abut 50~G and \vishal{show a reduced rise thereafter}. The intensities in the QS are larger than those in CH for the regions with identical {\bmag}. We further note that with increasing {\bmag}, and the difference in intensities increase slightly. This is similar to findings of~\cite{PradeepKashyap2018} for \ion{Mg}{2} lines and \cite{TriNS_2020} for \ion{Si}{4} line.
\subsection{Velocities} \label{sec:c2velocity}
\begin{figure*}[ht!]
\centering
\includegraphics[width=\linewidth]{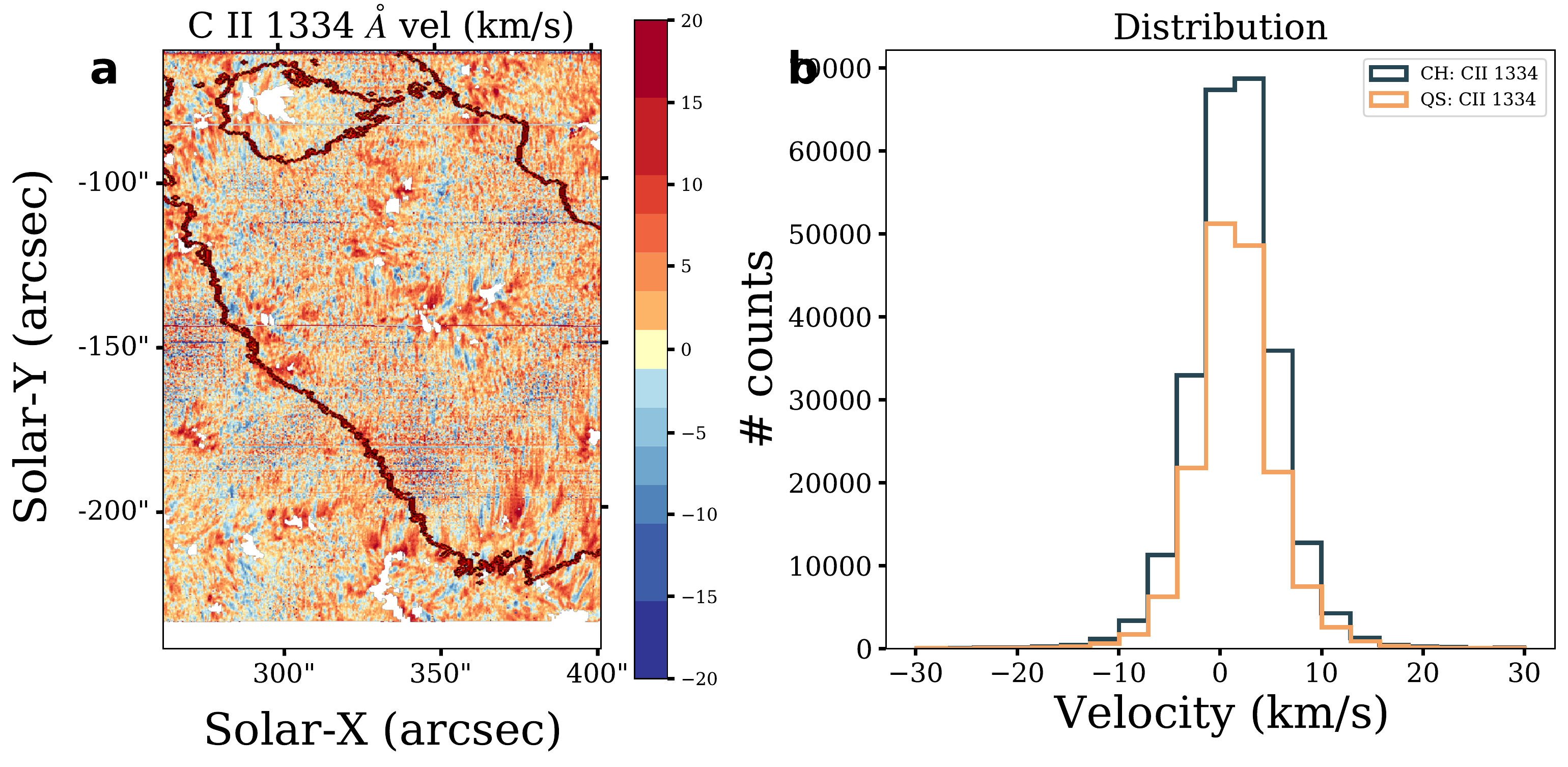}
  \caption{\vishal{Doppler map obtained in \car 1334~{\AA}  for DS3 is shown in panel {\bf a}.  The brown contours demarcate the boundary between the CH and QS. The velocity distribution in CH (black) and QS (orange) are shown in panel \textbf{b}.}}
  \label{fig:c2vel_map} 
\end{figure*}
\vishal{Fig.~\ref{fig:c2vel_map} displays the Doppler map obtained in \car 1334~{\AA} in panel {\bf a}, and the velocity distribution in panel \textbf{b}. The brown contours in panel {\bf a} demarcate CH and QS. The velocity maps shown in panel~\textbf{a} reveal that both the \car line is predominantly red shifted in CH as well as QS. Similar to the intensities, we find that there is no visual difference in the Doppler shift in the CH and QS. Black (orange) curves in panel {\bf b} denote CH(QS). The number of bins was again obtained using the rule of \cite{doanerule}. The histograms show marginally excess counts of both positive and negative velocities in CHs over QS in Fig.~\ref{fig:c2vel_map}.b. }

Similar to the intensities, we study the Doppler velocity both in QS and CH as a function of {\bmag}. For this purpose, we analyze this data in two ways. In the first method, we simply consider the average shift in every bin of {\bmag}. In the second, following \cite{TriNS_2020}, we consider the red-shifted and blue-shifted pixels separately for each bin of {\bmag}. The first method gives us the average flow for the CH and QS, while the second method gives us systematic variation of downflows and upflows with increasing {\bmag}. Such an exercise can reveal if the dynamics and structure of magnetic field causes any preferential effect on the Doppler shifts. 

\begin{figure}[ht!]
\centering
\includegraphics[width=0.95\textwidth]{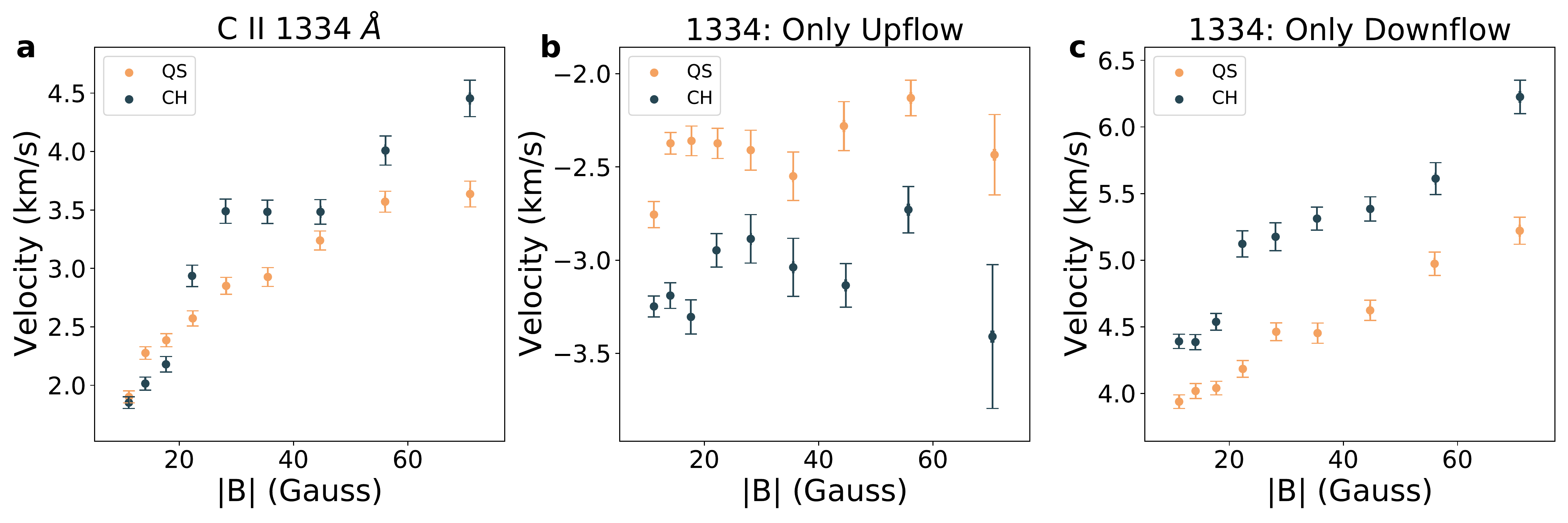}
\caption{Doppler shift as a function of {\bmag} for DS3. \vishal{Panel \textbf{a} shows the variation of average velocities obtained within bins of {\bmag}. Panel \textbf{b} shows the variation of velocities obtained for only blue-shifted pixels, while panel \textbf{c} shows the variation of only red-shifted pixels.}} 
\label{fig:c2_vel}
\end{figure}

Fig.~\ref{fig:c2_vel} displays the Doppler shifts as a function of {\bmag} measured in the CH and QS. \vishal{Panel {\bf a} shows the variation of signed average velocities obtained within bins of {\bmag}, while the panels {\bf b \& \bf c} depicts the variation of blueshifted and redshifted pixels alone, respectively. Panel~{\bf a} reveals that on average both QS and CH are red-shifted in chromosphere, and this velocity increases with {\bmag}}. When considering only the blue shifted pixels in Fig.~\ref{fig:c2_vel}.\textbf{b}, we find the CH have higher blueshifts than QS \vishal{, and that the blueshift appears independent of {\bmag} for QS, while being almost independent, within the uncertainties for CH.} Finally, when considering only the red-shifted pixels (see Fig.~\ref{fig:c2_vel}.\textbf{c}), we find that the CH have excess redshifts when compared to QS. We further note that the magnitude of the downflows is much larger than that of the upflows in both CH and QS, which explains the predominant downflows in the chromosphere. The excess downflows, however, \vishal{also} increase with increasing {\bmag}.

\subsection{Line width} \label{sec:c2width}
\begin{figure}[ht!]
\centering
\includegraphics[width=0.66\linewidth]{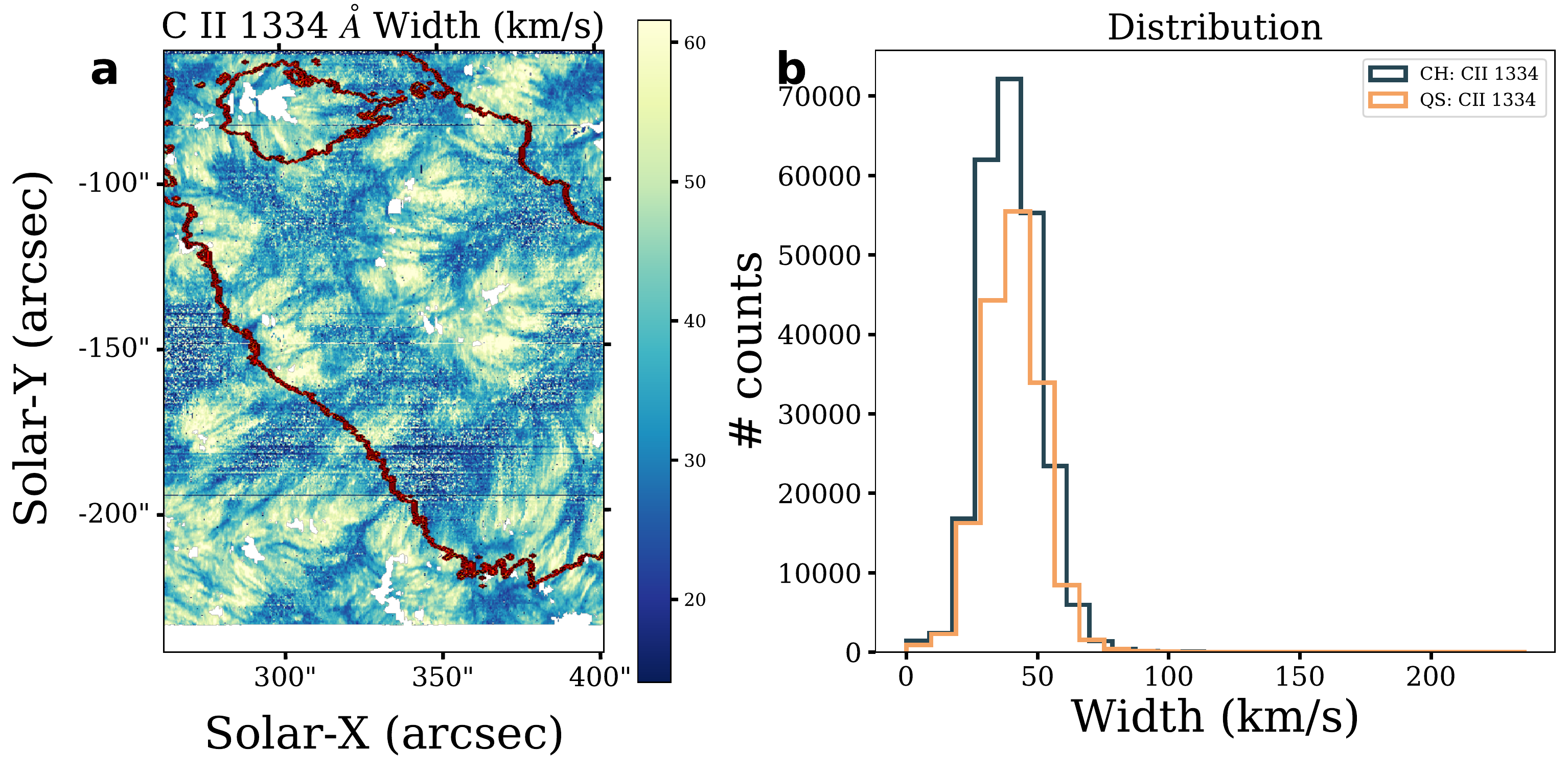}
\includegraphics[width=0.33\linewidth]{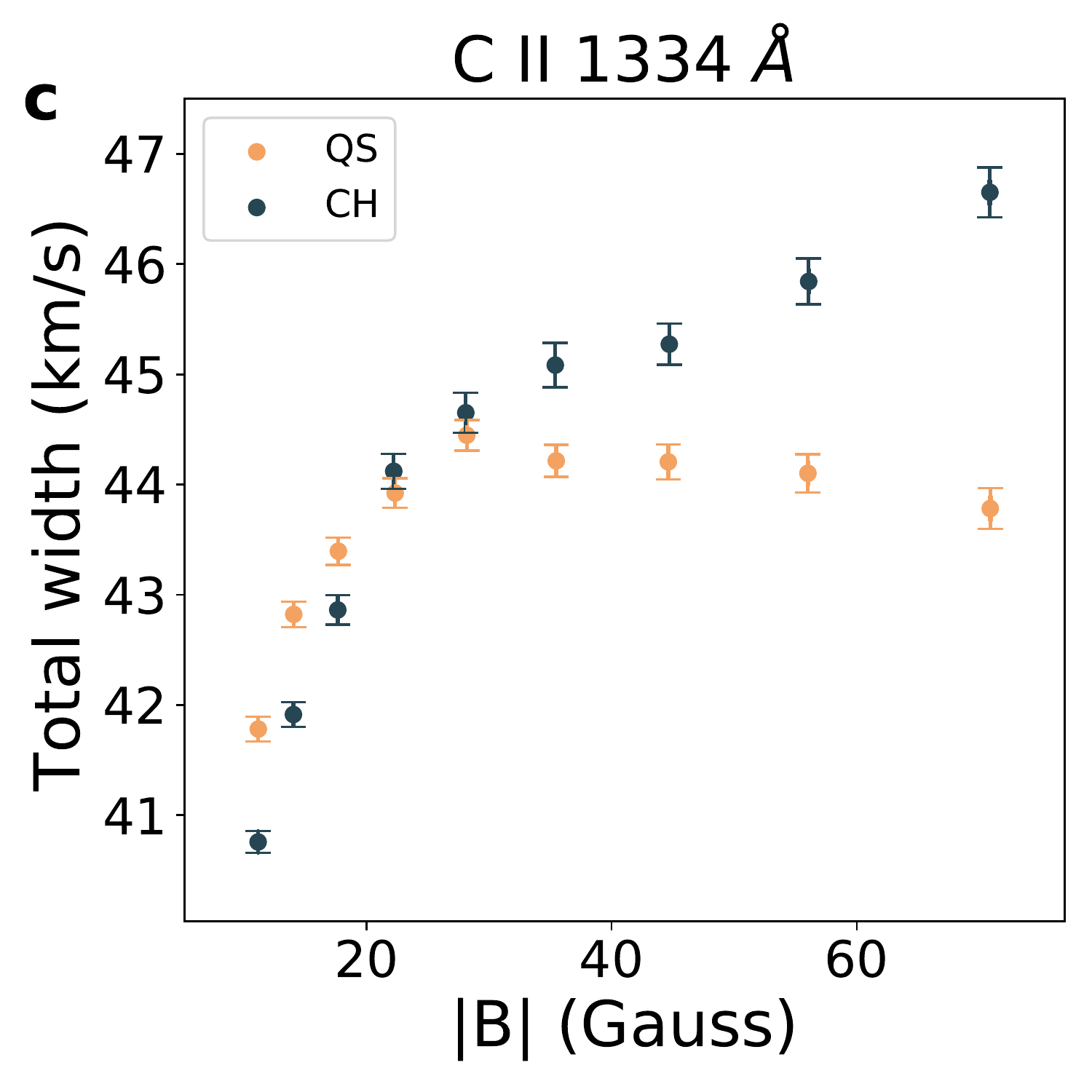}
  \caption{\vishal{Total line width map obtained from \car 1334~{\AA} for DS3 is shown in panel \textbf{a}. The brown contours demarcate the CH and QS. The line width distribution in CH (black) and QS (orange) are shown in panel \textbf{b}. The variation of line width with {\bmag} for CH (black) and QS (orange) is shown in panel \textbf{c} .}}
  \label{fig:c2w_map}
\end{figure}

We next study the total line width obtained from the Gaussian fit. Note that the line width \citep[see e.g.][]{Rathore_CII_paper2} is defined as:

\begin{equation}
  W_{\mathrm{FWHM}} = 2\sigma\sqrt{2\ln(2)}
\end{equation}

\noindent where $W_{\mathrm{FWHM}}$ is the line width, and $\sigma$ is standard deviation obtained from the Gaussian fits to the spectral line. 

\vishal{The line width map obtained for DS3 is shown in Fig.~\ref{fig:c2w_map}.a, with the brown contours demarcating the CH and QS. We plot the distribution of the width in panel \textbf{b}. Similar to the intensities and Doppler shifts, we do not see any conspicuous difference between the CH and QS. }

\vishal{In Fig.~\ref{fig:c2w_map}.c, we plot the line widths as a function of {\bmag}. Note that the bin size of the {\bmag} is same as those used for intensity and Doppler shift. The line width increases rapidly with increasing {\bmag}. Beyond 30{--}40~G, for CH, the width still increases, albeit slowly. However, for QS it shows saturation beyond 40~G and slight reduction thereafter. }

\subsection{Skew and Kurtosis} \label{sec:c2moment}
\begin{figure}[ht!]
  \centering
  \includegraphics[width=\linewidth]{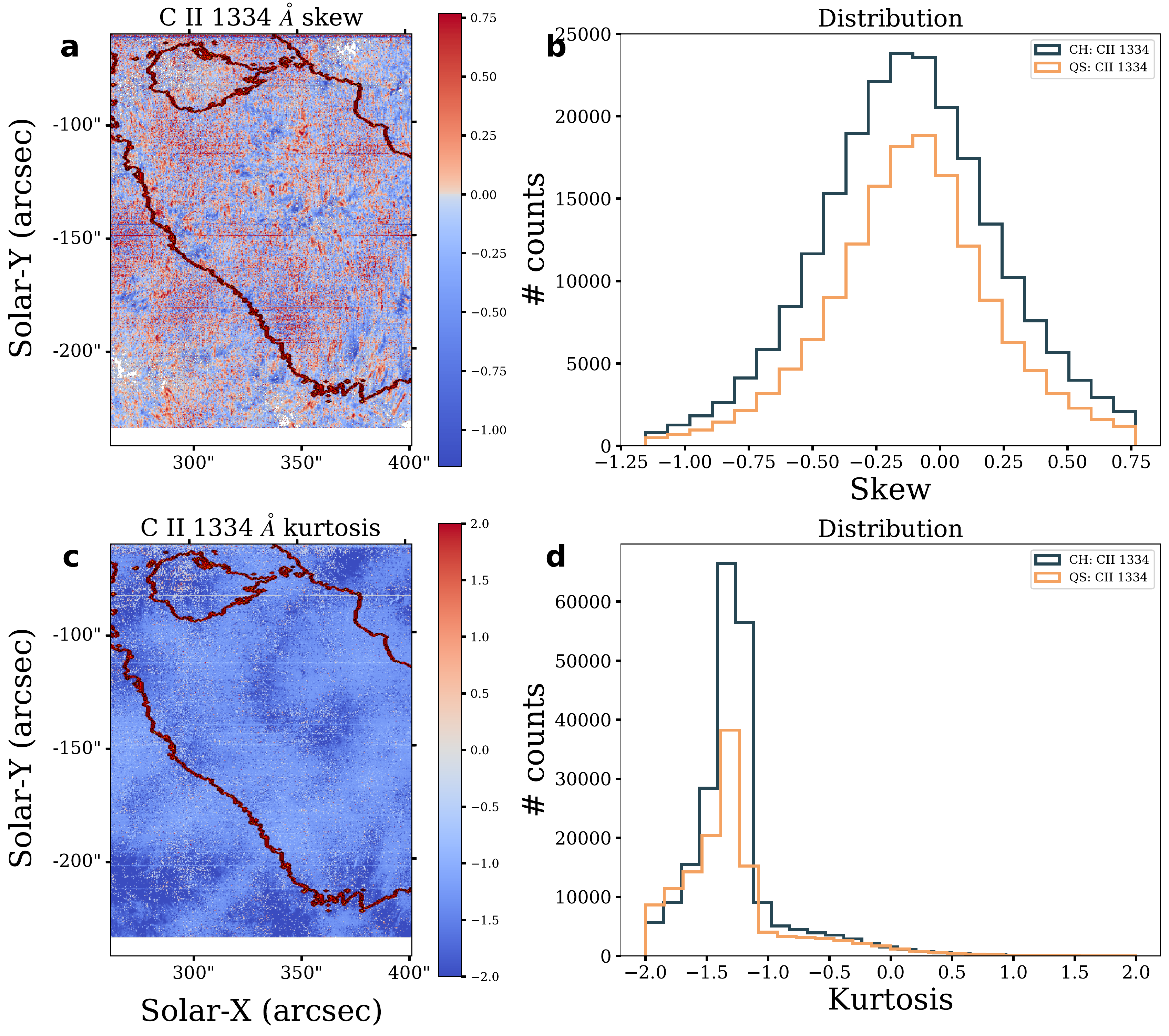}
  \caption{\vishal{Skew (panel \textbf{a}) and kurtosis (panel \textbf{c}) maps for \car 1334~{\AA} line obtained for DS3. The red contours depict the boundary between CH and QS. The distribution of skew and kurtosis in CH (black) and QS (orange) are shown in panel \textbf{b} and \textbf{d} respectively.}}
  \label{fig:c2mom_map}
\end{figure}
\begin{figure}[ht!]
\centering
\includegraphics[width=\linewidth]{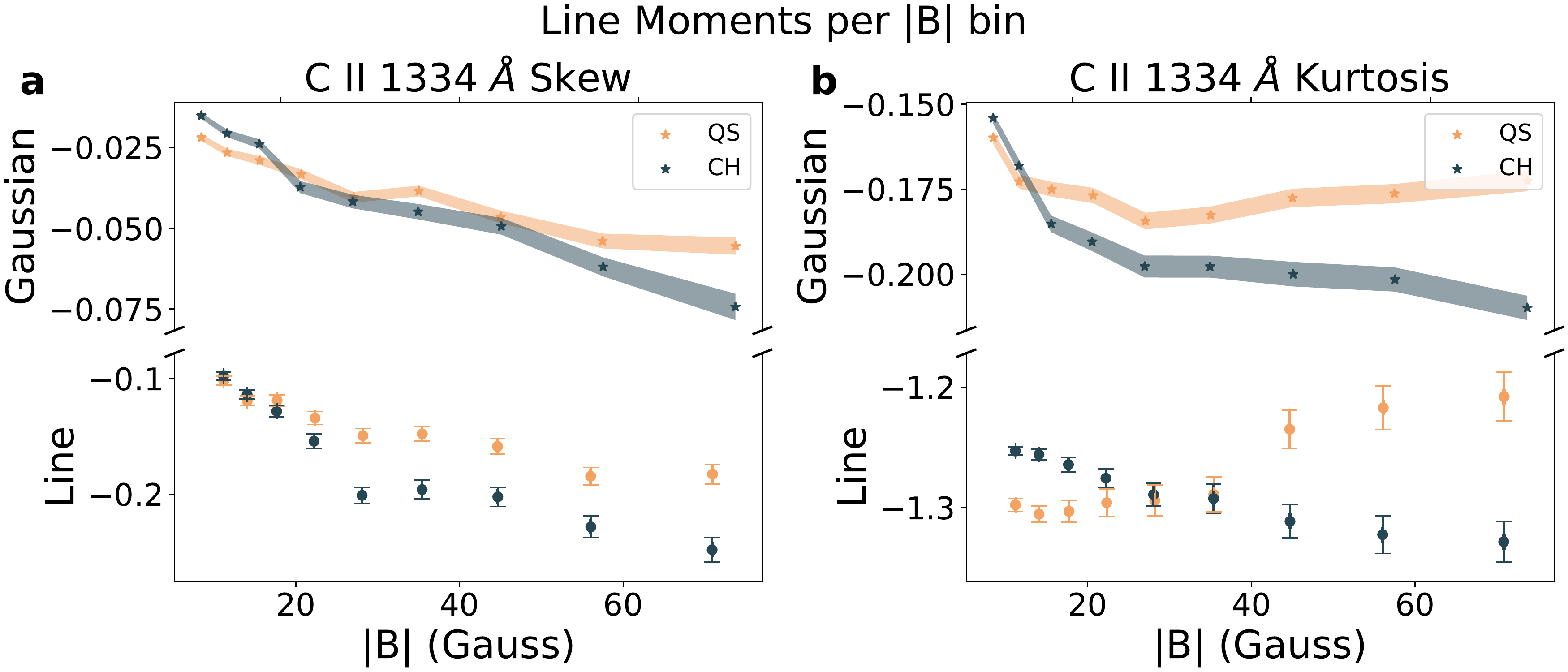}
  \caption{\vishal{Skew (panel \textbf{a}) and kurtosis (panel \textbf{b}) variation with {\bmag} for 1334~{\AA} for DS3.} The scatter plot with dots shows the moments for the spectral profiles, while the filled, star plots are computed for the single Gaussian fits over the same wavelength range. Note that the y-axis has been broken to show clearly the variation of moments with the {\bmag}.}
  \label{fig:c2moment}
\end{figure}

\vishal{Finally, we study the skew and kurtosis for the \car line using Eqn.~\ref{eqn:skew}~\&~\ref{eqn:kurt}. Fig.~\ref{fig:c2mom_map}.a (c) displays the skew (kurtosis) maps.} The skew maps show a good correspondence with that of magnetic field in Fig.~\ref{fig:segmentation_profiles}. This structure, however, is far more prominent as deficit of kurtosis \vishal{in Fig.~\ref{fig:c2mom_map}.\textbf{c}}. We plot the distribution of skew and kurtosis for CH (black) and QS (orange) \vishal{in Fig.~\ref{fig:c2mom_map}.b~\&~d respectively}. We find that the distribution of skew and kurtosis is very similar for both CHs and QS. 

In Fig.~\ref{fig:c2moment}, \vishal{we plot the variation of skew (panel a) and kurtosis (panel b) with {\bmag}.} In the plots, the skew and kurtosis obtained for the lines are shown as a scatter with dots, while the filled star bands correspond to the moments computed on the single Gaussian fit. \vishal{Theoretically, a Gaussian's skew and excess kurtosis are zero. However, this need not be the case for a Gaussian profile which is sampled at specific wavelength locations. Hence, to get an handle on the significance of the computed profile moments, we also compute the moments for the Gaussian fit as a benchmark. The deviations of obtained moments of the Gaussian profile from its theoretical values quantify the effects of a discrete wavelength grid.} To make the these clearer, the upper half of the plot are for moments obtained from Gaussian fits and lower half are for the \vishal{moments directly computed from the spectral profiles themselves}. The plots clearly reveal that the Gaussian fit and spectral line have significantly different moments. The spectral profiles are negatively skewed with respect to the Gaussian fits, indicating a general tendency to have a longer blue tail (or a steep red-ward rise) in the observed spectrum. Moreover, the line gets more skewed with increasing {\bmag}. Note that the errors are one sigma. The kurtosis plots in the right column show that the spectral lines are flatter, and have lesser outliers than a Gaussian due to kurtosis deficit. \vishal{The presence of significant differences indicate that these are not just due to sampling artifacts but also due to physical processes.}

\subsection{All datasets together}\label{sec:alldatasets}
\begin{figure}[ht!]
  \centering
\includegraphics[width=\linewidth]{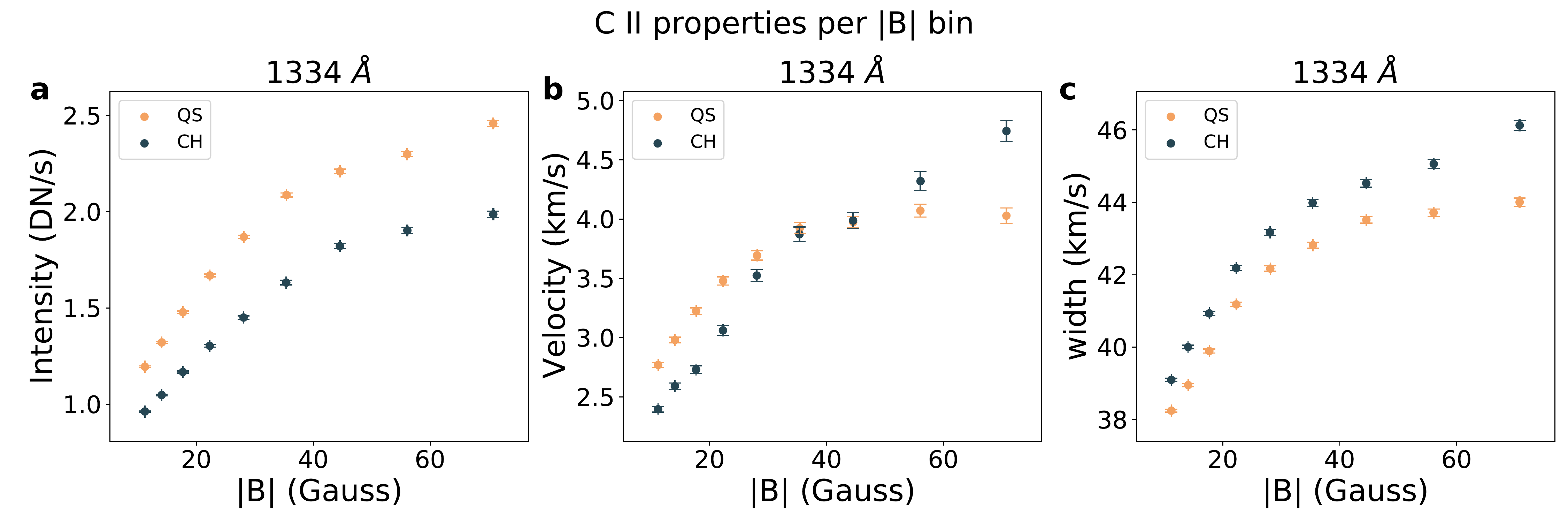}
  \caption{\vishal{\car intensity (panel \textbf{a}), velocity (panel \textbf{b}) and line width (panel \textbf{c}) variation with {\bmag} for all data sets taken together.}}
  \label{fig:c2_propeties_comb}
\end{figure}
\begin{figure}[ht!]
  \centering
  \includegraphics[width=\linewidth]{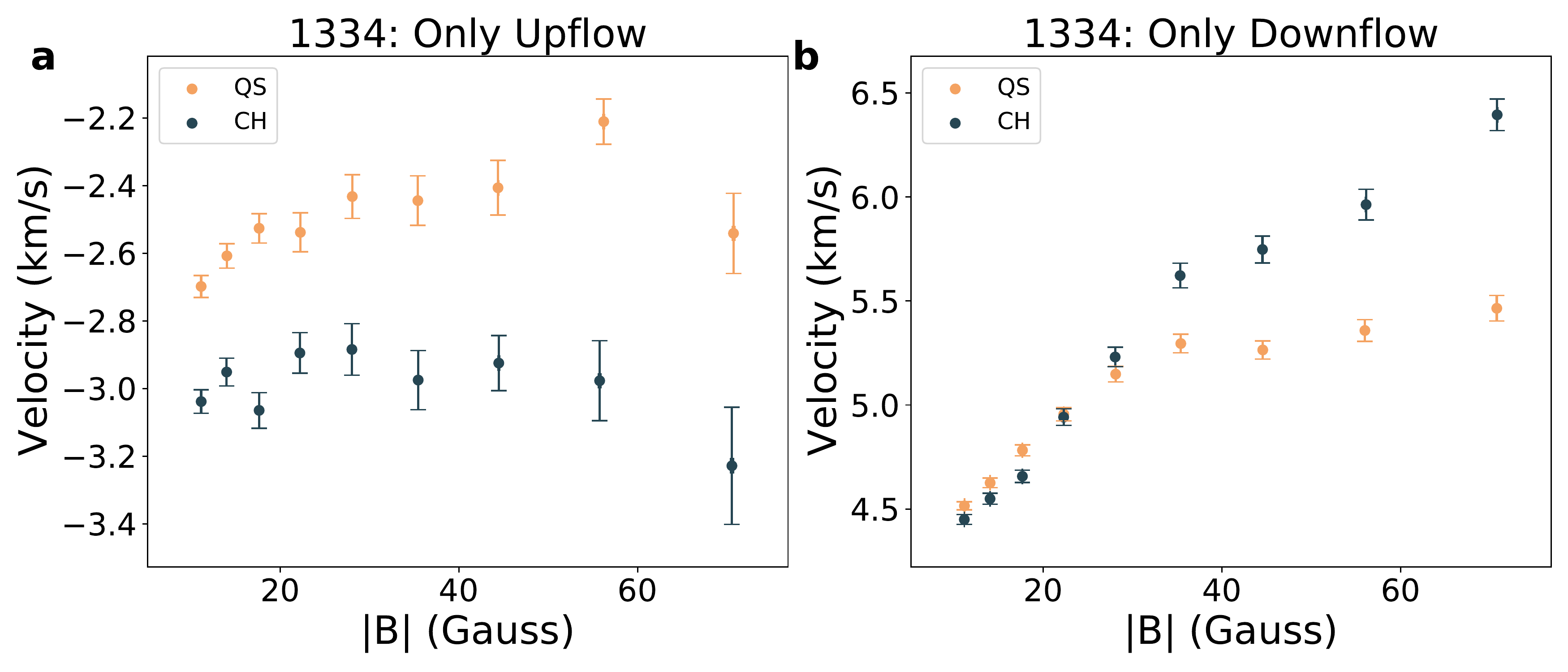}
  \caption{\car upflow and downflow velocity variation with {\bmag} for all data sets taken together. \vishal{Panel \textbf{a} shows the variation of upflows binned in {\bmag}, while panel \textbf{b} shows the variation for downflows. This plot is similar to Fig.~\ref{fig:c2_vel}, but is performed for all data sets taken together. }}
  \label{fig:c2_velocity_comb}
\end{figure}

As stated earlier, we analyze all the five dataset listed in Table.~\ref{tab:datadetails} and obtained similar results. To increases the statistical significance, we finally, average the obtained parameters from all five sets of observations, and study the dependence of intensity, velocity, line width, skew and kurtosis on {\bmag}. 

We display the results for intensity, velocity and width in Fig.~\ref{fig:c2_propeties_comb}. \vishal{ The plots reveal that the intensities increase in both QS and CHs as a function of {\bmag} (see panel a).} Moreover, QS regions have higher intensities than CHs for the regions with identical {\bmag}, and that the difference in the intensities increase with increasing {\bmag}. \vishal{The Doppler shifts plot (panel b) suggests} that both QS and CH are on average red-shifted and that the magnitude of the Doppler shift increases with increasing {\bmag}. We also note that for the smaller {\bmag} ($<$30~G), QS is slightly more redshifted than CH. Between 30{--}50~G, both show similar redshifts. At higher {\bmag} ($>$50~G), CHs are slightly more redshifted than QS. \vishal{Panel c shows} that the line width increases with {\bmag} and that CHs exhibit larger width than QS regions. 

In Fig.~\ref{fig:c2_velocity_comb}, we plot the velocity results for upflows and downflows separately as a function of {\bmag}. We find that the CH pixels are blue shifted relative to the QS pixels with identical {\bmag}. The blueshifts in CH show a marginal relation with the {\bmag}. Such a relation is not seen for QS, which in fact shows a \vishal{marginal reduction in blueshift with {\bmag} (see panel \textbf{a})}. \vishal{Figs.~\ref{fig:c2_velocity_comb}.\textbf{b} shows} that the redshifts in both CH and QS are almost same till $\approx30$ Gauss, following which the CHs show excess redshifts and QS shows saturation.

\begin{figure}[ht!]
  \centering
  \includegraphics[width=\linewidth]{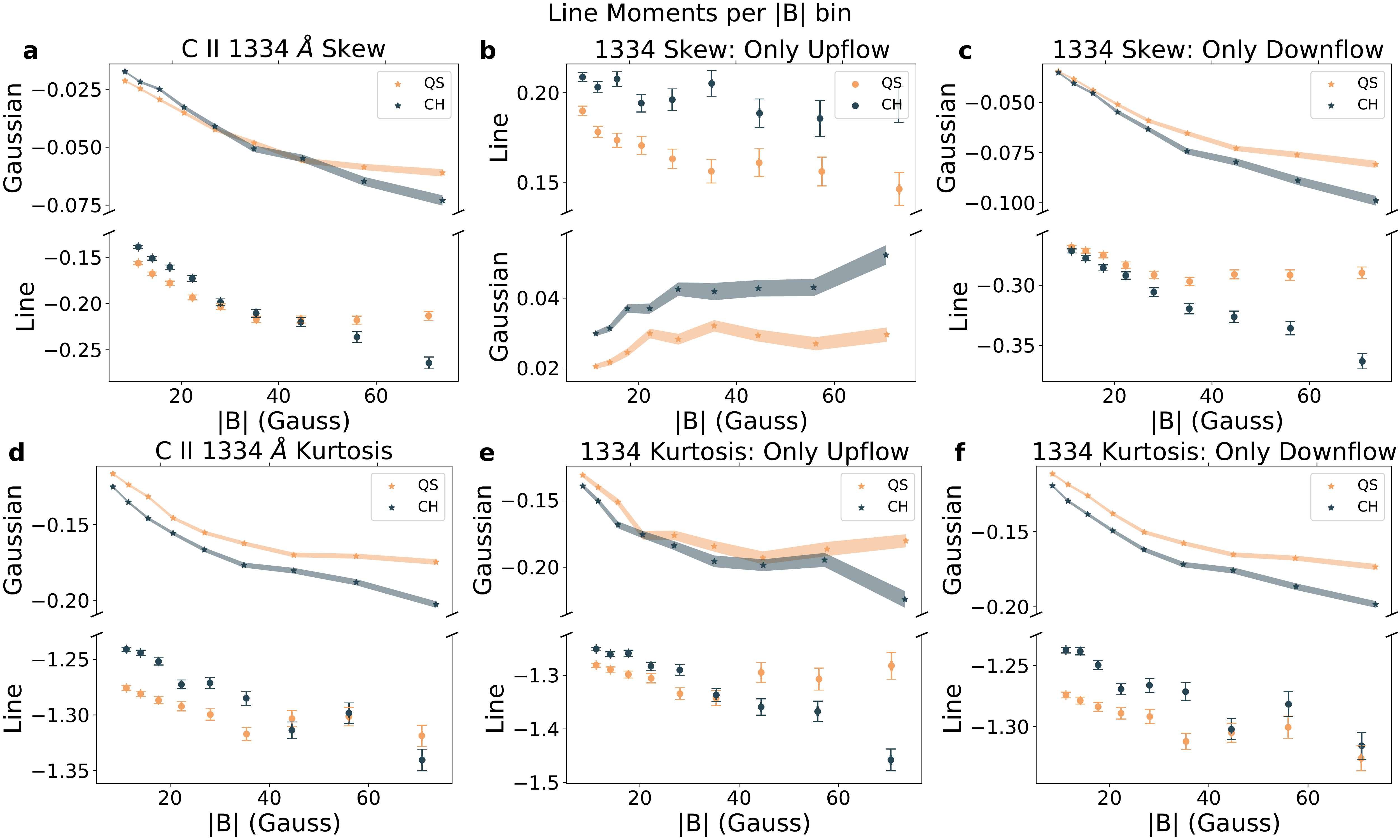}
  \caption{\car skew and kurtosis as a function {\bmag}, for all data sets taken together. The top row corresponds to line skew, while the bottom row corresponds to kurtosis. Panels \textbf{a} and \textbf{d} correspond to the variation of moments of all profiles, while \textbf{b} and \textbf{e} (\textbf{c} and \textbf{f}) correspond to moments of blueshifted (redshifted) profiles. The bands of black and orange, with stars over-plotted, correspond to the respective moment of a single Gaussian fit. The y-axis has been broken to depict the variation with {\bmag} better.}
  \label{fig:c2_moments_comb}
\end{figure}

Finally, in Fig.~\ref{fig:c2_moments_comb}, we plot the skew (panel {\bf a}) and kurtosis (panel {\bf d}) averaged over the five sets of observations as a function of {\bmag}. \vishal{Similar to Fig.\ref{fig:c2moment}}, the star and banded plots depict the moments for the Gaussian fit, while the dots depict moments for the spectral profile. We find a clear signal of kurtosis deficit and negative skew of the lines vis-a-vis a Gaussian profile. Furthermore, we also study the moments for red- and blue shifted pixels separately \vishal{(see panel \textbf{b} and \textbf{c})}. We find that blueshifted (redshifted) pixels are positively (negatively) skewed. Such a behavior suggests that the spectra with blue (red) shifts rise more steeply than a Gaussian on the blueward (redward) side, and fall of gradually on the opposite side. Finally, the kurtosis shows no dependence on the line shift, as seen from panels \textbf{e} and \textbf{f} which have kurtosis as a function of redshift and blueshift of the line. This implies that the spectral profiles themselves are flatter than a Gaussian profile irrespective of whether they are shifted to blue or red side.

\section{Summary, discussion and conclusions}\label{sec:discuss}

\vishal{The \car resonance line observed at around 1334~{\AA} provides us with extensive information on the plasma conditions in the solar chromosphere. Specifically, the intensity, Doppler shift and width are standard quantities that may be inferred from this line.} Moreover, the line skew and kurtosis may provide us with information on the variation of the source function in the chromosphere \citep[see e.g.,][]{Rathore_CII_paper1,Rathore_CII_paper2}. In this work, we study and compare the properties of the \vishal{\car line} in CH and QS regions as a function of {\bmag}. For this purpose, we have used the observations recorded by IRIS, HMI and AIA. The results are presented for a single dataset from Fig.~\ref{fig:c2i_map} through Fig.~\ref{fig:c2moment}, while the results obtained by combining  all datasets are shown in Figs.~\ref{fig:c2_propeties_comb} through Fig.~\ref{fig:c2_moments_comb}. Below, we summarize the obtained results.

\begin{enumerate}
  \item CH have lower intensities vis-\`a-vis QS for the regions with identical {\bmag}, and this difference increases with increasing {\bmag}.
  
  \item Both CH and QS are, on average, redshifted in the \vishal{\car line}. The QS shows larger redshifts for {\bmag} $\lessapprox30$ Gauss, while the CHs show larger redshifts for {\bmag} $\gtrapprox50$ Gauss. The QS velocities, however, show a saturation at large {\bmag}, while the CH velocities show a monotonic increase with {\bmag}.
  
  \item For pixels with only upflows, CHs have larger blueshifts vis-\`a-vis QS. This blueshift marginally increases with increasing {\bmag} for CH. The QS blueshifts, however, remain constant (or even marginally reduce) with increasing {\bmag}.
  
  \item For pixels with only downflows, both CHs and QS show similar redshifts till $\approx30$ Gauss. At higher {\bmag}, the CHs show excess redshift over QS. We also note that while the redshifts in CHs keep increasing monotonically, those in QS shows saturation beyond 40~G.
  
  \item The total line width, i.e., the FWHM in both QS and CH show monotonic increase with increasing {\bmag}, with some sign of saturation beyond 50~G. CHs have excess total line widths vis-\`a-vis QS for the regions with identical {\bmag}, and this difference slightly increases with increasing {\bmag}, albeit at the level of saturation.
  
  \item Both CH and QS spectral profiles have similar skew and kurtosis. However, the profiles themselves are negatively skewed and flatter vis-\`a-vis a Gaussian profile. The regions with larger {\bmag} clearly show a kurtosis deficit and an average negative skew. 
  
\end{enumerate}

A detailed comparison between various parameters observed in CHs and QS have been performed in various lines, spanning a broad range of temperatures using the observations recorded by the Solar Ultraviolet Measurements of Emitted Radiation \citep[SUMER;][]{sumer,Stucki_UVLinesSumer_Chs, Stucki_UVLinesSumer_correlations,Stucki_FUVLinesSumer,xia_chsumer} and Coronal Diagnostic Spectrometer \citep[CDS;][]{cds, StuSP_2002}, both on board the Solar and Heliospheric Observatory (SoHO), and IRIS \citep[e.g.][]{PradeepKashyap2018,TriNS_2020}. \vishal{At chromosphere heights}, by studying the \ion{N}{1} (1319~{\AA}, $\approx1.6\times10^4$ K) and \ion{Ni}{2} (1317~{\AA}, $\approx1.4\times10^4$ K), \cite{Stucki_FUVLinesSumer} found that CHs have marginally higher intensities over QS. However, as also mentioned by the authors, this difference could be due to an artifact of the statistics. For the \mg line, which also forms in the chromosphere, \cite{PradeepKashyap2018} found that CHs have reduced intensities over QS regions, for the regions with identical {\bmag}. These two results are different from each other. However, it is important to emphasize here that \cite{PradeepKashyap2018} compares the intensities in CHs and QS for the regions with identical {\bmag} and not the simple area-averages, as was done by \cite{Stucki_FUVLinesSumer}. Our results for \vishal{the \car line agree} with those obtained by \cite{PradeepKashyap2018}, albeit \car generally forms a bit higher than \mg line as shown by \cite{Rathore_CII_paper1}. Note that the intensity ratio of QS to CH at $\approx70$ Gauss is $\approx1.2$ for \ion{Mg}{2} peak \citep[][]{PradeepKashyap2018}, $\approx1.42$ in \car and $\approx1.6$ for \ion{Si}{4}~\citep[][]{TriNS_2020} . This is in-line with increasing difference between CH and QS from chromosphere to the corona, with increasing height.


The observed differences in \mg intensities by \cite{PradeepKashyap2018} have been attributed to loop statistics proposed by \cite{Weigelmann_loopstats}, which is obtained based on scaling laws. The intensity differences in CH and QS, in this scenario, arise simply due to the presence of excess (similar) long (short) closed loops in QS vis-\`a-vis the CHs. Similar interpretation can be provided here too. However, as also noted by \cite{PradeepKashyap2018}, such scaling laws may not be directly applicable in for the observations derived using spectral lines formed in chromospheric region. Finally, note that using 3D MHD simulations \cite{Rathore_CII_paper1,Rathore_CII_paper2} have shown that the intensities of the \car lines depends on the source function at $\tau=1$ height. Our results hence suggest that the values of the source function at line formation height \vishal{are smaller} in CH over QS.

With SUMER observations, \cite{Stucki_FUVLinesSumer,Stucki_UVLinesSumer_Chs,xia_chsumer} reported that CHs have marginally excess redshifts over QS in \car 1334{~\AA}, albeit within the uncertainties. These results are, again, different than those observed here using IRIS observations. Our results conspicuously show that for the regions with {\bmag} $<$~($>$) 40 (50)~G, QS (CH) is clearly more redshifted than CHs (QS). We further note that while the flows in QS saturates at higher flux values i.e. for network regions, those in CH show a monotonic increase. \cite{He2008_SWModelling} performed a 1-D hydrodynamic simulation of a cylindrically symmetric flux tube with impulsive deposition of energy at $\approx5$~Mm height, following the suggestions by \cite{tu2005solar} and demonstrated the presence of upflows (downflows) at heights above (below) 5Mm. Such an impulsive energy deposition scenario can potentially explain the presence of excess redshifts in CH over QS. However, it is important to note that the energy deposition may occur across a range of heights as observed in the simulations by \cite{Hansteen2010_SWModelling}. Thus there may be upflowing plasma at \car temperatures if energy is dumped at much lower heights. Such scenario can potentially explain the observed excess upflows in CH over QS. However, this needs further investigation, combining observations and simulations.

From simulations, it has been observed that opacity plays an important role in the broadening of the lines \citep{Rathore_CII_paper1,Rathore_CII_paper2}. Opacity broadening may be qualitatively explained by Eq.25 of \citep{Rathore_CII_paper1}, where in the absence of any flows the opacity broadening is proportional to the ratio of column mass at the line wing to the column mass as line center. From Eq. 23 and 20 of~\citep{Rathore_CII_paper1}, we find:
\begin{equation}
  \frac{1}{m_c(0)} = \frac{\chi_{l0}}{\rho} + \frac{1}{m_c(\infty)},
  \label{eqn:colmass}
\end{equation}
\noindent where $m_c(\Delta\nu)$ is the column mass at a shift $\Delta\nu$, $\Delta\nu=0$ representing the line core, $\Delta\nu=\infty$ representing the continuum, $\chi_{l0}$ the opacity at line core per unit volume, and $\rho$ being the density. Thus, with other terms being constant, $m_c(0)$ depends directly on the density $\rho$, and any reduction in density reduces the line core column mass, thereby increasing the opacity broadening. Assuming the line intensity to be directly related to density, a reduction in density would be seen as a reduction in the core intensity. Thus, density reduction in the line core of CH over QS can neatly explain the observed intensity and line width differences. However, note that Eq.~\ref{eqn:colmass} has been derived under a static atmosphere, while in the real solar atmosphere there would be components of non thermal velocities and micro-turbulence that will affect the line width.

\vishal{Our observations further show that the spectral profiles themselves have} less kurtosis than a Gaussian, and are negatively skewed vis-\`a-vis a Gaussian profile. To understand these profiles further, we look at the skew and kurtosis of redshifted and blueshifted profiles separately, and attempt to disentangle their properties in Fig.~\ref{fig:c2_moments_comb}. The skew is observed to change sign depending on the line shift. The observed profiles are observed to be positively (negatively) skewed if the profile is blueshifted (redshifted). Since the comparison is performed with respect to a Gaussian fit, it would mean that the blueshifted (redshifted) profiles have a steeper blueward (redward) rise than a Gaussian. Such asymmetric \car profiles have been observed in 1-D simulations by \cite{Avrett_2013_1DC2}. Moreover, the authors have observed increasing asymmetry with increasing atmospheric velocities. It has been suggested by \cite{Avrett_2013_1DC2} that the asymmetry arises if the flows are column mass conserving -- implying that the vertical velocity is inversely proportional to the density. Hence, the part of line that is emitted higher shows greater shift than the part of line emitted lower. This may be a possible explanation for the observed skew of the line. 


Finally, we clearly see that the kurtosis is independent of whether the profile is blueshifted or redshifted, and is significantly different from a Gaussian. It also shows a distinct variation with {\bmag}. Thus, \car profiles are, in general, flatter than a Gaussian, and the flatness increases with increasing {\bmag}. The \ion{Ca}{2} lines in spicules have been shown to change from having a central reversal to a flat topped to a peaked profile with increasing formation height by \cite{Zirker_ca2}. Such changes in profiles were explained by a reduction in opacity in these lines. Similar picture may also hold with the \car line, which may show such flat topped profiles due to opacity variations. Note, however, that similar kurtosis-deficit profiles have been clearly seen as the presence of a ``box-shaped'' profile by \cite{Rathore_CII_paper2}. From 3D simulations, \cite{Rathore_CII_paper2} assert this to be a consequence of a steep rise in the source function near the continuum, with a more gradual rise near the core formation region. Also note that such a source function variation would also give rise to broader lines, as shown in \cite{Rathore_CII_paper2}. Thus, the flat rise of  source function, dictated by the underlying {\bmag}, may cause the kurtosis deficit in the \car line.

The results presented in this paper demonstrate the diagnostics potential of \vishal{the \car line} and provide further important inputs for modeling of the solar chromosphere. Further observational work in synergy with simulations is definitely warranted.

\vishal{We sincerely thank the referee for their insightful comments and suggestions.} We thank Prof. Mats Carlsson (University of Oslo) for various discussions. \vishal{This work is partially supported by the Max-Planck Partner Group of MPS at IUCAA.}. We acknowledge the use of data from IRIS, AIA and HMI. IRIS is a NASA small explorer mission developed and operated by LMSAL with mission operations executed at NASA Ames Research Center and major contributions to downlink communications funded by ESA and the Norwegian Space Centre. AIA and HMI are instruments on board SDO, a mission for NASA's Living With a Star program.

\software{Numpy~\citep{numpy_nature},  Astropy~\citep{astropy1}, Sunpy~\citep{sunpy}, Scipy~\citep{scipy},  Scikit-image~\citep{scikit-image}, Matplotlib~\citep{matplotlib}, Multiprocessing~\citep{multiprocessing}, OpenCV~\citep{opencv_library}. Jupyter~\citep{jupyter}.}

\bibliography{manuscript}{}
\bibliographystyle{aasjournal}



\end{document}